\documentstyle[seceq,epsfig]{ptptex}

\title{Kramers-Wannier Approximation for the 3D Ising Model}

\author
{ Kouichi {\sc Okunishi}\footnote{E-mail  address:
okunishi@godzilla.phys.sci.osaka-u.ac.jp} and
Tomotoshi {\sc Nishino}$^{1}$\footnote{E-mail address:
nishino@phys.sci.kobe-u.ac.jp}
}

\inst
{Department of Applied Physics, Graduate school of Engineering,
Osaka University, Suita, Osaka 565-0781\\
$^1$Department of Physics, Graduate School of Science,
Kobe University, Inoshishidai 657-8501}

\recdate{}

\abst{
We investigate the Kramers-Wannier approximation for the
three-dimensional (3D) Ising model. The variational state is represented by
an effective 2D Ising model, which contains two variational parameters. We
numerically calculate the variational partition function using the corner
transfer matrix renormalization group (CTMRG) method, and find its maximum
with respect to the variational parameters. The calculated transition point
$K_{\rm c} = 0.2184$ is only 1.5\% less than the true $K_{\rm c}$; the result
is better than that obtained by the corner transfer tensor renormalization
group (CTTRG) approach. The calculated phase transition
is mean-field like.
}

\begin{document}

\maketitle

\section{Introduction}

In 1941 Kramers and Wannier~\cite{Krm} proposed a variational approximation
for the two-dimensional (2D) Ising model, which is called `Kramers-Wannier
(KW) approximation' today. The feature of the approximation is that the
variational state is constructed as the thermal equilibrium state of the 1D
Ising model in an effective magnetic field. From the modern view point, their
variational state can be regarded as an example of so called the matrix
product state.~\cite{AKLT,Fannes} More than 20 years later, Baxter improved
the KW approximation by introducing additional degrees of freedom into the
variational state; he reformulated the variational principle of the KW
approximation using the corner transfer matrix (CTM).~\cite{Bx1,Bx2,Bx3} It
has been known that the variational property in both the KW approximation and
Baxter's CTM formulation has many aspects in common with that in the density
matrix renormalization group (DMRG)~\cite{Wh1,Wh2,Ostlund,rev} and the
recurrent variational ansatz.~\cite{rev,RVA1}

The transition temperature and the specific heat of the 2D Ising model
calculated by the KW approximation are more accurate than those obtained by
the mean-field approximation and the Bethe approximation.~\cite{Bethe} It is
expected that the KW approximation is a good non-perturbative method also in
higher dimensions. We therefore investigate the KW approximation  for the 3D
Ising model by way of the maximization of the Rayleigh ratio
\begin{equation}
\lambda = \frac{\langle V | T | V \rangle}{\langle V | V \rangle}  \, ,
\label{rratio}
\end{equation}
where $T$ is the `layer-to-layer' transfer matrix, and $| V \rangle$ is the
variational state represented as a 2D generalization of the matrix product
state. Though the variational formulation is quite simple, such a
generalization of the KW approximation to 3D systems has not been
investigated so far. This is partially because there is no
analytical tool to calculate $\langle V | T | V \rangle$ and $\langle V | V
\rangle$ in eq.~(\ref{rratio}) that are partition functions of unsolvable 2D
lattice models. In this paper we overcome the problem by numerically
calculating $\lambda$ using the corner transfer matrix renormalization group
(CTMRG),~\cite{CTMRG1,CTMRG2} which is a variant of the DMRG for 2D classical
systems~\cite{rev,Ni,Ni2,Carlon1} formulated via  Baxter's
CTM.~\cite{Bx1,Bx2,Bx3}
Our approach shown in the following can be regarded as the DMRG applied to 3D classical systems.
It should be noted that the formulation of the KW approximation in eq.~(\ref{rratio}) is related to the tensor product variational formulation for 2D quantum
systems.~\cite{Zitt2,Hieida,Sierra}

In the next section, we introduce the KW variational state $| V \rangle$
for the 3D Ising model, and present the concrete definition of
eq.~(\ref{rratio}). In \S 3, we explain the way to apply CTMRG to the
variational formulation, and then show the calculated spontaneous
magnetization and the internal energy. Conclusions are summarized in \S 4,
and we discuss several possible improvements to the formulation of the KW
approximation in 3D.

\section{Variational Formulation in 3D}

We consider the 3D Ising model on a simple cubic lattice of the size $N
\times N \times L$ in the $X$, $Y$, and $Z$ direction, respectively, where on
each lattice point $(i,j,k)$ --- the position $(i,j)$ in the $k$-th spin
layer --- there is an Ising spin $\sigma_{i\,j}^k = \pm 1$. We assume open
boundary conditions in both $X$ and $Y$ directions, and periodic boundary
condition in $Z$-direction. The Hamiltonian of the 3D Ising model is
\begin{equation}
H = - J \sum_{ijk} (  \sigma_{i'\!j}^k \sigma_{i\,j}^k
+ \sigma_{i\,j'}^k \sigma_{i\,j}^k + \sigma_{i\,j}^{k'} \sigma_{i\,j}^k ) \, ,
\end{equation}
where we have used the notation $i' \equiv i + 1$, $j' \equiv j + 1$, and
$k' \equiv k+1$ for book keeping. Throughout this paper we consider the
ferromagnetic case $J>0$. The partition function of the system is expressed as
\begin{equation}
Z = \sum_{\{\sigma\}} {\rm exp}( - \beta H ) = {\rm Tr} \, T^L_{~} \, ,
\end{equation}
where $T$ is the layer-to-layer transfer matrix, and the sum is taken over
for all the spin configurations. In the following, we consider the
symmetrized transfer matrix $T (= T^T_{~})$ constructed as a  product of
local Boltzmann weights
\begin{equation}
T( \sigma^{k'} | \, \sigma^k ) = \, \prod_{ij} \, W_{i\,j}^{k} \, ,
\end{equation}
where $\sigma^k$ and $\sigma^{k'}$ represent spin configurations in $k$-th
and $k$+$1$-th layer, respectively,~\cite{drop} and  $W_{i\,j}^k$ is the local
Boltzmann weight defined by
\begin{eqnarray}
W_{i\,j}^{k}
= {\rm exp} \biggl\{ \frac{K}{4}
&(&
\sigma_{i'\!j}^{k'} \sigma_{i\,j}^{k'} + \sigma_{i\,j'}^{k'}
\sigma_{i\,j}^{k'} +
\sigma_{i'\!j'}^{k'} \sigma_{i'\!j}^{k'} +\sigma_{i'\!j'}^{k'}
\sigma_{i\,j'}^{k'}
\biggr. \nonumber\\
\biggl.
&+&
\sigma_{i\,j}^{k'} \sigma_{i\,j}^k + \sigma_{i'\!j}^{k'} \sigma_{i'\!j}^k +
\sigma_{i\,j'}^{k'} \sigma_{i\,j'}^k + \sigma_{i'\!j'}^{k'} \sigma_{i'\!j'}^k
\biggr. \nonumber\\
\biggl.
&+&
\sigma_{i'\!j}^k \sigma_{i\,j}^k + \sigma_{i\,j'}^k \sigma_{i\,j}^k +
\sigma_{i'\!j'}^k \sigma_{i'\!j}^k + \sigma_{i'\!j'}^k \sigma_{i\,j'}^k )
\biggr\} \, , \label{isingweight}
\end{eqnarray}
with  $K \equiv \beta J$. In the r.h.s of the eq.~(\ref{isingweight}),  the
12 terms correspond to the 12 edges of a local cube,  and the coefficient
$1/4$ of $K$ denotes that each bond between the nearest neighbor spins is shared by 4 adjacent cubes.

As an introduction to the KW approximation for the 3D Ising model, let us
consider a special mean-field approximation, which replaces all the Ising
spins $\sigma_{i\,j}^k$ except at the $k$-th spin layer $\sigma^k$ by their
expectation value $\langle\sigma\rangle$. The approximation draws the
effective Hamiltonian for the $k$-th spin layer
\begin{equation}
{\bar H}(\sigma^k)
= - J \sum_{ij} (  \sigma_{i'\!j}^k \sigma_{i\,j}^k
+ \sigma_{i\,j'}^k \sigma_{i\,j}^k + 2\langle\sigma\rangle \sigma_{i\,j}^k
) \, ,
\end{equation}
which is nothing but the Hamiltonian of the 2D Ising model under the mean
field $2J\langle\sigma\rangle$ imposed from both the up and down sides of the
$k$-th layer. In this mean-field framework, the spin profile in the $k$-th
layer is given by the weight $P(\sigma^k) = \exp\{-\beta  \bar{H}(\sigma^k)
\}$. It is expected that the mean field weight $P(\sigma^k)$ well
approximates the appearance probability of the layer-spin configuration, and
that its square root
\begin{eqnarray}
\label{paramet}
\sqrt{P(\sigma^k)}
&=&
\prod_{ij} {\rm exp} \biggl\{ \frac{K\langle\sigma\rangle}{4} (
\sigma_{i\,j}^k +\sigma_{i'\!j}^k + \sigma_{i\,j'}^k +\sigma_{i'\!j'}^k )
\biggr.
\nonumber\\
&&
~~~~~~~~~~
\biggl. + \, \frac{K}{4} (
\sigma_{i'\!j}^k \sigma_{i\,j}^k + \sigma_{i\,j'}^k \sigma_{i\,j}^k +
\sigma_{i'\!j'}^k \sigma_{i'\!j}^k + \sigma_{i'\!j'}^k \sigma_{i\,j'}^k )
\biggr\}  \, ,
\end{eqnarray}
which is proportional to $T( \sigma^{k} | \langle\sigma\rangle )$, can be
used for the variational state in eq.~(\ref{rratio}).

Such a direct usage of $\sqrt{P(\sigma^k)}$ as the variational state,
however, has a shortcoming in the paramagnetic region, where
$\langle\sigma\rangle$ is zero and $\sqrt{P(\sigma^k)} $ has no adjustable
parameter. Following Kramers and Wannier, we introduce an additional
parameter to the nearest neighbor coupling term in eq.~(\ref{paramet}). The
variational state (in the product form) is then given by
\begin{eqnarray}
\label{variats}
V( \sigma^k ) = \prod_{ij} U_{i\,j}^k \, ,
\label{vs}
\end{eqnarray}
where the local factor $ U_{i\,j}^k$ is defined as
\begin{eqnarray}
 U_{i\,j}^k  &\equiv& \exp \left\{ \frac{h}{4} (
\sigma_{i\,j}^k + \sigma_{i'\!j}^k + \sigma_{i\,j'}^k + \sigma_{i'\!j'}^k )
\right.
\nonumber \\
&{}& ~~~  + \left. \frac{g}{4} (
\sigma_{i'\!j}^k \sigma_{i\,j}^k + \sigma_{i\,j'}^k \sigma_{i\,j}^k +
\sigma_{i'\!j'}^k \sigma_{i'\!j}^k + \sigma_{i'\!j'}^k \sigma_{i\,j'}^k )
 \right\}  \, ,
\label{localvs}
\end{eqnarray}
with two variational parameters $h$ ($=$ effective magnetic field) and
$g$ ($=$ effective nearest neighbor coupling). The variational state
$V(\sigma^k )$ in eq.~(\ref{variats}) has at least one variational parameter
$g$ even when the system is paramagnetic ($h = 0$). Substituting $T(
\sigma^{k'} | \, \sigma^k )$ and $V( \sigma^k )$ to the variational
formulation in eq.~(\ref{rratio}), the Rayleigh ratio --- the approximation
partition function  per layer of the size $N \times N$ --- is expressed
as
\begin{equation}
\lambda_N = \frac{ \sum_{ \{\sigma^{k'}\} \{\sigma^k\} }
V( \sigma^{k'} ) \, T( \sigma^{k'} | \, \sigma^k ) \, V( \sigma^k )
}{ \sum_{ \{\sigma^k\} } V( \sigma^k ) V( \sigma^k )} \, ,
\end{equation}
where the denominator of the r.h.s.
\begin{eqnarray}
A_N = \sum_{\{\sigma^k\}} \prod_{ij}  \left( U_{i\,j}^k \right)^2
&=&
\sum_{\{\sigma^k\}}\prod_{ij} {\rm exp} \biggl\{ \frac{h}{2} (
\sigma_{i\,j}^k +\sigma_{i'\!j}^k + \sigma_{i\,j'}^k +\sigma_{i'\!j'}^k
) \biggr. \\
&&
~~~~~~~~~~~~~~
\biggl. + \, \frac{g}{2} (
\sigma_{i'\!j}^k \sigma_{i\,j}^k + \sigma_{i\,j'}^k \sigma_{i\,j}^k +
\sigma_{i'\!j'}^k \sigma_{i'\!j}^k + \sigma_{i'\!j'}^k \sigma_{i\,j'}^k )
\biggr\}    \nonumber
\end{eqnarray}
is a partition function of an effective 2D Ising model parameterized by $h$
and $g$.
Similarly, the numerator
\begin{equation}
B_N \equiv  \sum_{\{\sigma^{k'}\} \{\sigma^k\}}
V( \sigma^{k'} ) \, T( \sigma^{k'} | \, \sigma^k ) \, V( \sigma^k )
= \sum_{\{\sigma^{k'}\} \{\sigma^k\}}\prod_{ij}
U_{i\,j}^k W_{i\,j}^{k} U_{i\,j}^{k'}
\end{equation}
is a partition function of a two-layer Ising model parameterized by $h$,
$g$, and $K$.

\section{Numerical Result}

The goal of the KW approximation is to find out the pair of $h$ and $g$ ---
as functions of $K$ --- that maximizes the variational partition function per
site in the thermodynamic limit:
\begin{equation}
z(K,h,g)
= \lim_{N\rightarrow\infty} (\lambda_N)^{1/N^2}
= \lim_{N\rightarrow\infty} (B_N / A_N)^{1/N^2} \, .
\label{tag3-1}
\end{equation}
In order to find out the maximum of $z(K,h,g)$ in the $h$-$g$ parameter
space, we calculate $z(K,h,g)$ for various values of $h$ and $g$ via the
numerical calculation of $A_N$ and $B_N$ for $N = 3, 5, 7,\ldots$ up to a
sufficiently large $N$. After that we search the maximum of $z(K,h,g)$ in the
$h$-$g$ plane. From the numerical point of view, it is better to use the
formulation
\begin{equation}
z(K,h,g) = \lim_{N\rightarrow\infty}
\left(\frac{B_{N+4}A_{N+2}B_N}{A_{N+4}B_{N+2}A_N}\right)^{1/8}
\end{equation}
in order to accelerate the numerical convergence with respect to $N$,
rather than just taking the limit $N \rightarrow \infty$ directly to $(B_N /
A_N)^{1/N^2}$ .

We use the CTMRG method~\cite{CTMRG1,CTMRG2} for the calculation of $A_N$
and $B_N$, since the method enables us to obtain $A_N$ and $B_N$ very
rapidly and accurately. We keep $m = 32$ states in the CTMRG calculations,
and obtain $A_N$ and $B_N$ up to $N = 150$; the condition is sufficient for
the precise determination of $z(K,g,h)$. Throughout this section we set $J =
1$ and thus $K = \beta$.

\begin{figure}[htb]
\parbox{\halftext}{
\epsfxsize=52mm
\centerline{\epsffile{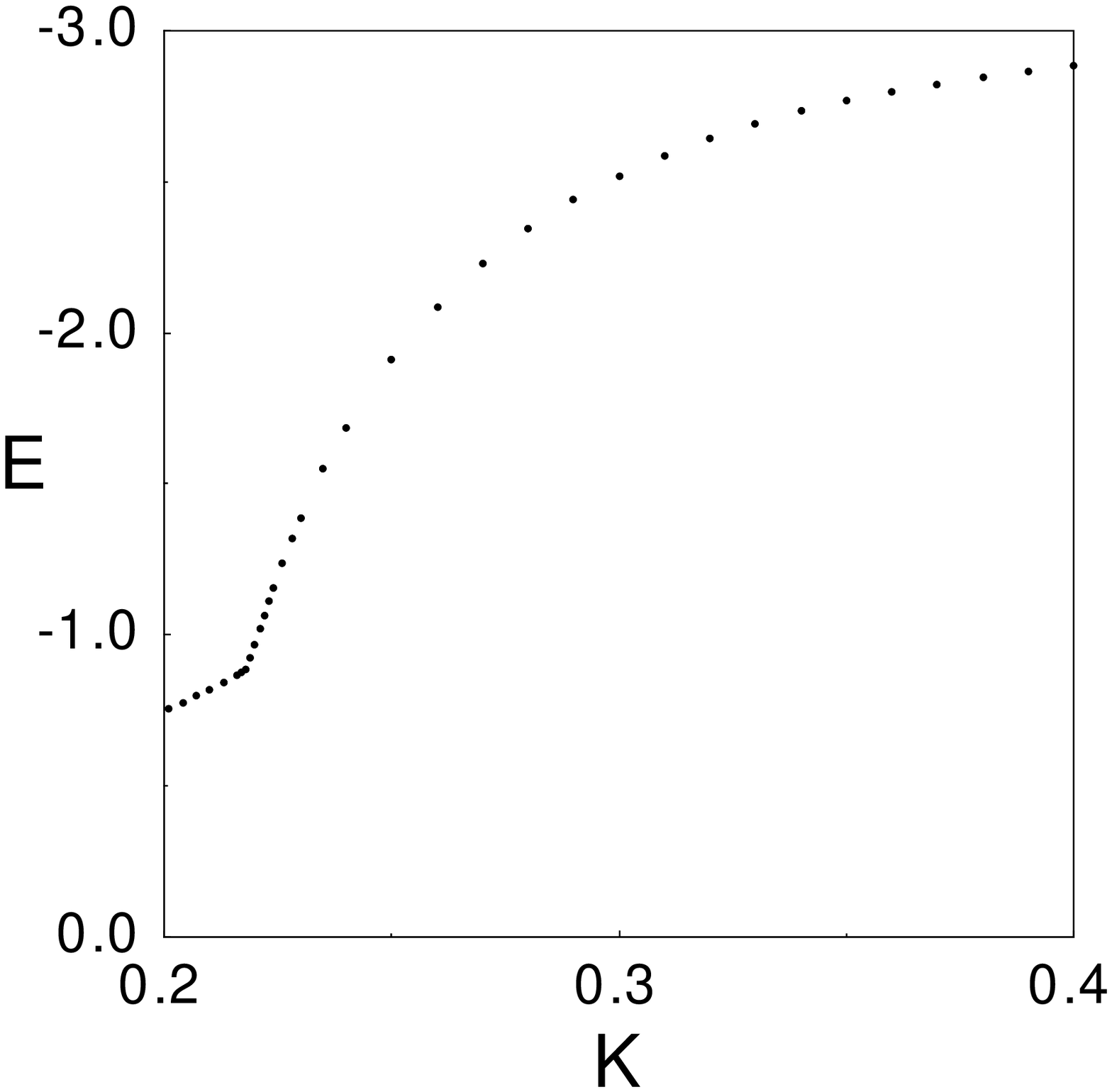}}
\caption{ The internal energy $E$ as a function of $K$.
The transition point is observed at $K_{\rm c} = 0.2184$. }
\label{fig1}
}
\hspace{8mm}
\parbox{\halftext}{
\epsfxsize=52mm
\centerline{\epsffile{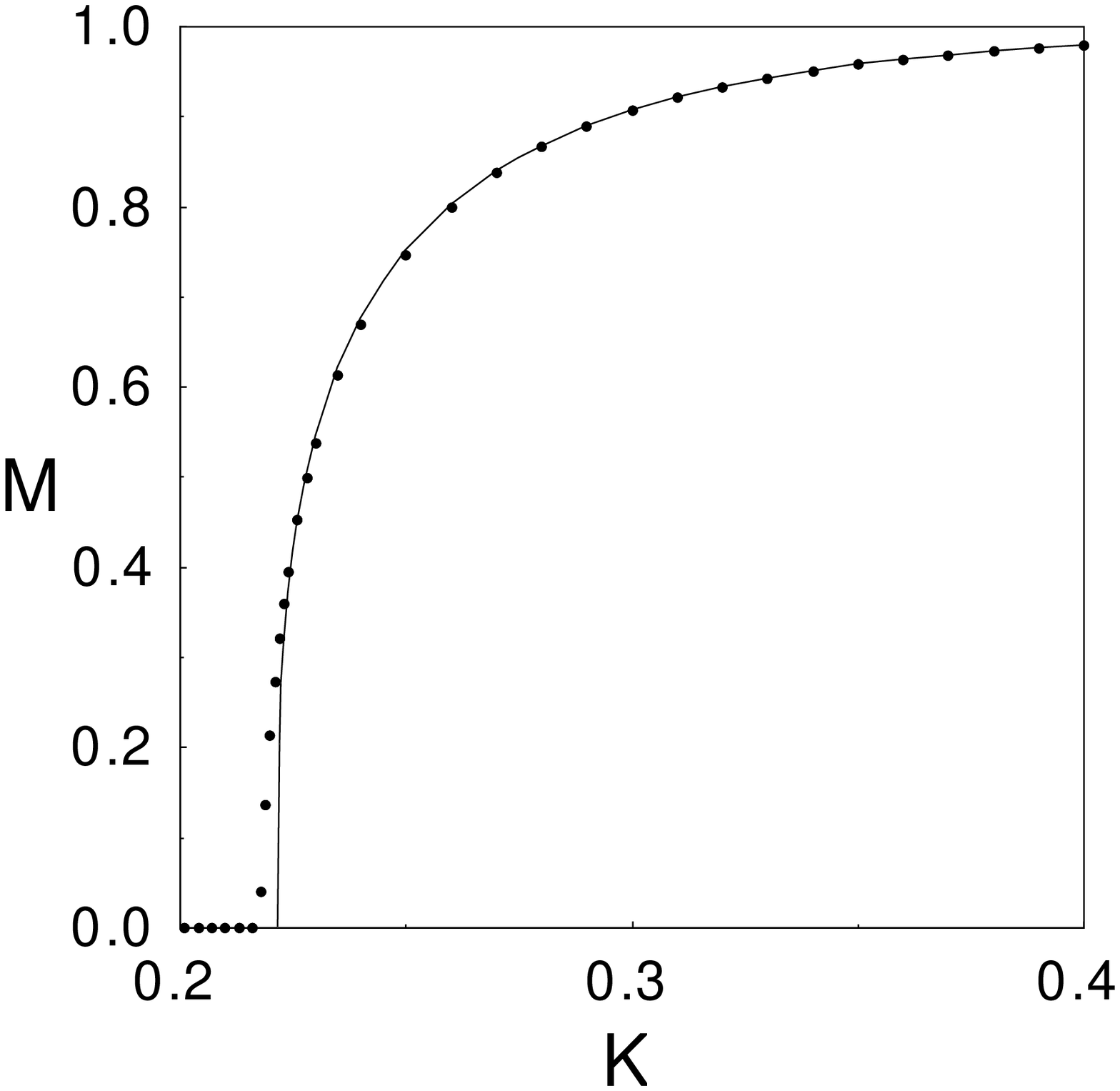}}
\caption{ The magnetization $M$ as a function of $K$. The solid line denotes
eq.(\ref{correctiontoscaling}) obtained by Monte Carlo
simulations.~\cite{MC1,MC2}  In
the very vicinity of the calculated transition point, $M$ is
proportional to $\sqrt{K -K_{\rm c}}$.} \label{fig2} }
\end{figure}

First, let us see the calculated results of thermodynamic quantities. Figure~\ref{fig1} shows the internal energy per cube, which is given by
\begin{equation}
E = - \langle \sigma_{i\,j}^k \sigma_{i'\,j}^{k}\rangle - \langle
\sigma_{i\,j}^k\sigma_{i\,j'}^k\rangle -
\langle\sigma_{i\,j}^k\sigma_{i\,j}^{k'}\rangle \, ,
\label{eng}
\end{equation}
according to the formal thermodynamic relation $E = -\frac{\partial}{\partial\beta} \ln z(K,h,g)$.
We have checked that the numerically calculated $\ln z(K,h,g)$ really satisfies the relation.
The plotted data in Fig.~\ref{fig1} has a kink at $K_{\rm c} = 0.2184$, which is the  transition point from the paramagnetic state to the ferromagnetic state. The calculated $K_{\rm c}$ is
about 1.5\% smaller than one of the reliable critical point $K_{\rm c}^{\rm
MC} = 0.2216544 \pm 0.000005$ determined by Monte Carlo
simulations.~\cite{MC1,MC2} The discrepancy 1.5\% shows that the KW
approximation for the 3D Ising model is more accurate  than that for the 2D
Ising model; for the latter, the critical point $K_{\rm c} = 0.4122$
calculated by the KW approximation is 6.5\% smaller than the exact one
$K_{\rm c} = 0.4407$.~\cite{Onsager}

In Fig.~\ref{fig2}, we draw the spontaneous magnetization $M \equiv
\langle\sigma_{i\,j}^k \rangle$. For comparison, we also show Tarpov and
Bl\"ote's  Monte Carlo result for the cubic lattice containing up to $256^3$ spins, (see eq. (10) in Ref.[22])
\begin{eqnarray}
M^{\rm MC}
= t^{0.32694109}(1.6919045-0.34357731t^{0.50842026} -042572366t ) \, , \,\,
\label{correctiontoscaling}
\end{eqnarray}
where $t\equiv 1-K_{\rm c}^{\rm MC}/K$, and this expresson is very accurate in $t<0.26$.
The KW results agrees with eq.~(\ref{correctiontoscaling}) in almost whole the region of $K$.
In the very vicinity of the transition point $K_{\rm c} < K < K_{\rm c} +0.01$,
the calculated magnetization deviates from eq.~(\ref{correctiontoscaling}) and
behaves as $5.171 \sqrt{K - K_{\rm c}}$ approximately. In principle, the phase
transition observed by the KW approximation is mean-field like.~\cite{Surda}

\begin{figure}[htb]
\parbox{\halftext}{
\epsfxsize=52mm
\centerline{\epsffile{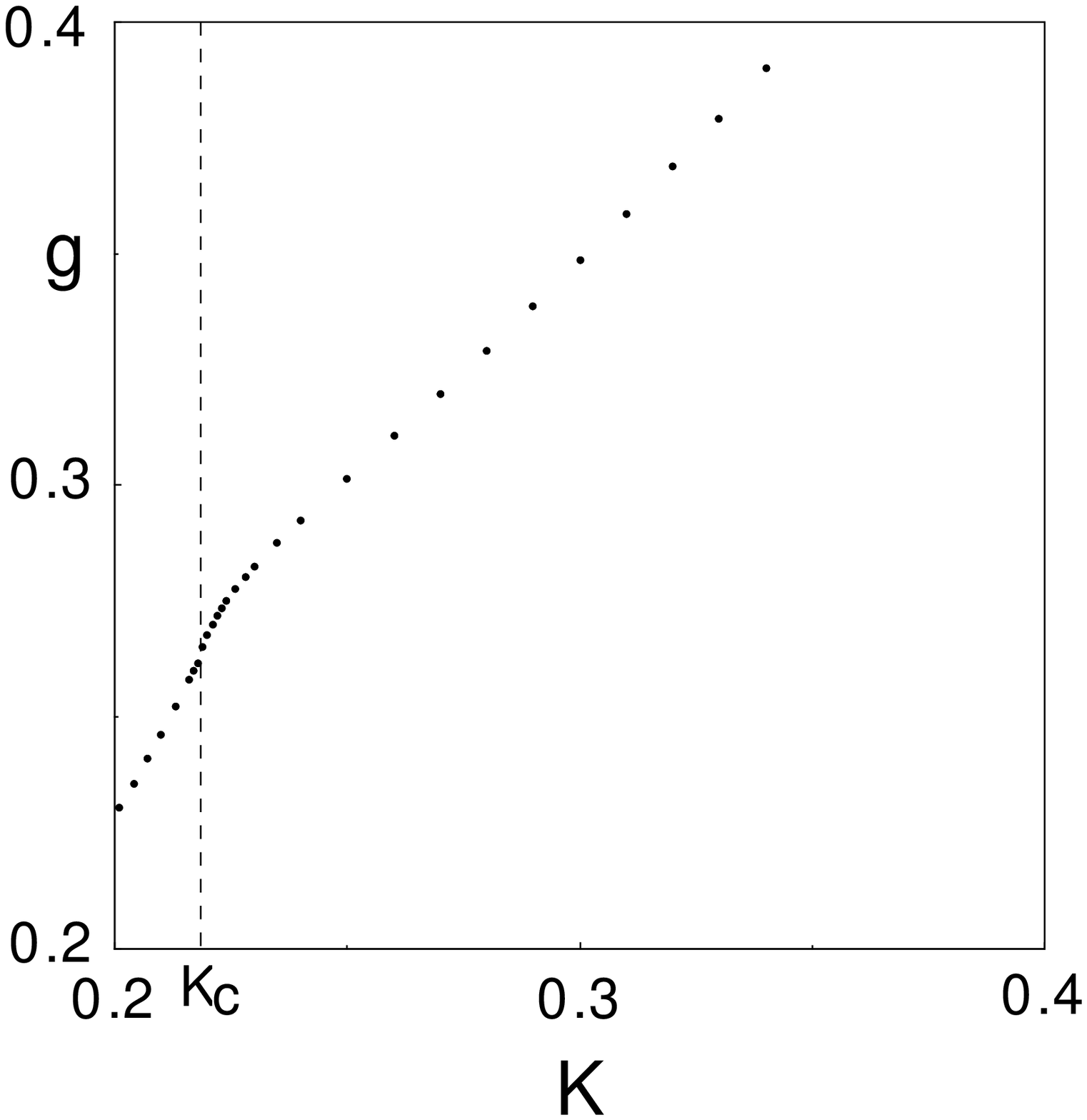}}
\caption{ The variational parameter $g$ as a function of $K$. }
\label{fig3}
}
\hspace{8mm}
\parbox{\halftext}{
\epsfxsize=52mm
\centerline{\epsffile{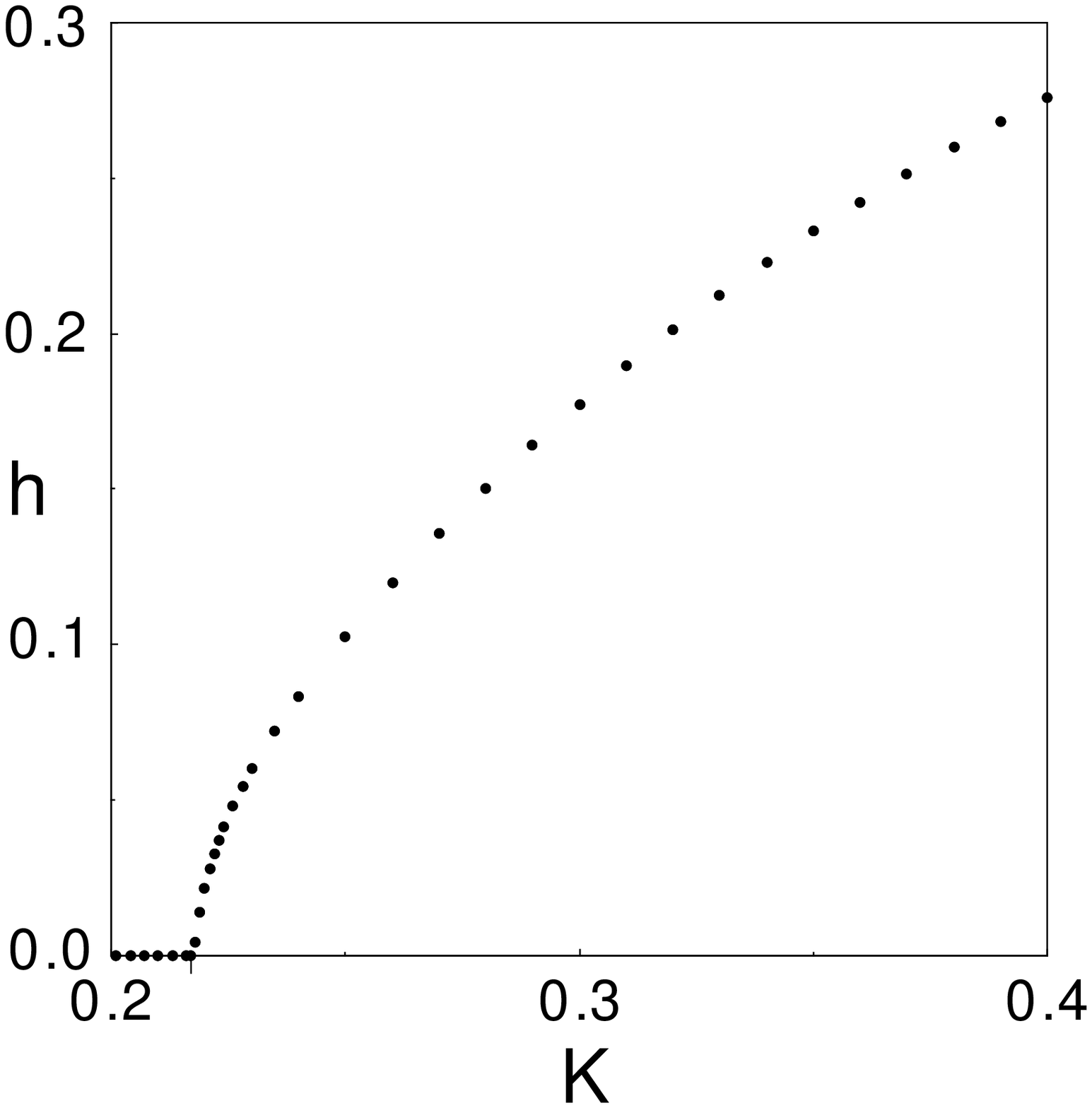}}
\caption{ The variational parameter $h$ as a function of $K$.
In the vicinity of the transition point, $h$ is proportional to
$\sqrt{K - K_{\rm c}}$.}
\label{fig4}
}
\end{figure}

Let us see the properties of the variational parameters $g$ and $h$,
respectively, since they are closely related to $E$ and
$M$. As the internal energy $E$ in Fig.~\ref{fig1}, the
parameter $g$ shown in Fig.~\ref{fig3} has  a kink at the calculated $K_{\rm
c}$, and is always larger than $K$. The parameter $h$ shown in
Fig.~\ref{fig4} is approximately $0.5516 \sqrt{K - K_{\rm c}}$ in  the
vicinity of the calculated $K_{\rm c}$; $2h \simeq KM$ is
approximately  satisfied in the neighborhood of $K_{\rm c}$.

\section{Conclusion and Discussion}

We have applied the KW approximation to the 3D Ising model, representing the variational state as the thermal equilibrium state of an effective 2D Ising model.
We have calculated the variational partition function numerically using CTMRG method, and maximized the function with respect to the variational parameters $g$ and $h$.

The KW approximation draws the spontaneous magnetization fairly well in wide
region of temperature, compared with the Monte Carlo simulations.
The calculated  transition point $K_{\rm c} = 0.2184$ is only 1.5\%
smaller than one of the most reliable $K_{\rm c}$ determined by Monte Carlo
simulations;~\cite{MC1,MC2}  the $K_{\rm c}$ obtained in the KW approximation
is better than that obtained by the corner tensor renormalization group
(CTTRG).~\cite{CTTRG} The critical behavior observed by the KW approximation
is mean-field like in the very vicinity of the transition point.

There are at least two ways to improve the variational state used in the KW
approximation for the 3D Ising model. A way is to introduce additional
variational parameters into the trial state $V(\sigma^k)$. Thermal equilibrium
state of arbitrary 2D classical lattice models, such as the multi-layer 2D
Ising model and the Ising model with next nearest neighbor interactions, can
be the candidates for $V(\sigma^k)$. It is straightforward to apply the CTMRG to such a variational state to evaluate the variational partition function.
However, the number of variational parameters is limited by the numerical effort to find out optimal variational parameter sets.
The other way is to introduce block spin variables into each local factors in $V(\sigma^k)$.~\cite{Sierra,Ni2}
Although this approach contains much more variational parameters than the former way, we can  treat the optimization problem more systematically as was done in the CTTRG.\cite{CTTRG}
In both of these improvements, the key point is to find out the best variational parameter set quickly.

We finally comment that the formulation of the KW approximation for the 3D
Ising model presented here can be applied to various 2D quantum spin
systems. The generalization is simply to replace the 2D tensor product
state in Hieida's DMRG formulation~\cite{Hieida} by the thermal equilibrium
state of 2D classical lattice models. 
This is not a trivial simplification, since the relation between the 2D KW variational state and the 2D tensor product state have not been clarified yet, unlike the trivial relation in 1D.

\section*{Acknowledgments}

The authors thank to Y.~Akutsu and Y.~Hieida and N. Maeshima for valuable
discussions. The  present work is partially supported by a Grant-in-Aid from
Ministry of Education, Science and  Culture of Japan. K.~O. is supported by
Japan Society for the Promotion of Science.

\end{document}